\documentclass[12pt]{article}
\usepackage{amsmath,amssymb,epsfig}
\begin{document}
\parindent=0pt
\parskip=6pt
\rm

\begin{center}
{\bf \large Magnetic effects on the phase transitions in\\ unconventional
superconductors}

{\bf Dimo I. Uzunov}

 Max-Plank-Institut  f\"{u}r Physik komplexer
Systeme, N\"{o}tnitzer Str. 38, 01187 Dresden, Germany,

[Permanent address:  CP Laboratory, Institute of Solid
State Physics, Bulgarian Academy of Sciences, BG-1784 Sofia,
Bulgaria.]

\end{center}

\begin{abstract}
 The effect of magnetic fluctuations on the critical behaviour of unconventional
ferromagnetic superconductors (UGe$_2$, URhGe, etc.) and
superfluids is investigated by the renormalization-group method.
For the case of isotropic ferromagnetic order a new unusual
critical behaviour is predicted. It is also shown that the
uniaxial and bi-axial magnetic symmetries produce fluctuation
driven phase transitions of first order. The results can be used
in interpretations of experimental data and for a further
development of the theory of critical phenomena in complex
systems.
\end{abstract}

Experiments~\cite{Saxena:2000, Huxley:2001} at low temperatures
($T \sim 1$ K) and high pressure ($P\sim 1$ GPa) demonstrated the
existence of spin triplet superconducting states in the metallic
compound UGe$_2$. The superconductivity is triggered by the
spontaneous magnetization of the ferromagnetic phase that occurs
at much higher temperatures~\cite{ShopovaPL:2003,Shopova:2005,
Shopova:2006}. The ferromagnetic order coexists with the
superconducting phase in the whole domain of its existence below
$T \sim 1.2$. The same phenomenon of existence of
superconductivity at low temperatures and high pressure in the
domain of the $(T,P)$ phase diagram where the ferromagnetic order
is present was observed in URhGe~\cite{Aoki:2001}). These
remarkable phenomena occur through phase transitions of first and
second order and multi-critical points which present a
considerable experimental and theoretical interest. A fragment of
($P,T$) phase diagrams of itinerant ferromagnetic
compounds~\cite{Saxena:2000} is sketched in fig.~1, where the
lines $T_F(P)$ and $T_c(P)$ of the
paramagnetic(P)-to-ferromagnetic(F) and
ferromagnetic-to-coexistence phase(C) transitions are very close
to each other and intersect at very low temperature or terminate
at the absolute zero ($P_0,0$). At low temperature, where the
phase transition lines are close enough to each other, the
interaction between the real magnetization vector
$\mbox{\boldmath$M$}(\mbox{\boldmath$r$)} =
\{M_j(\mbox{\boldmath$r$}); j=1,...,m \}$ and the complex order
parameter vector of the spin-triplet Cooper
pairing~\cite{Sigrist:1991}, $\psi(\mbox{\boldmath$r$}) =
\{\psi_{\alpha}(\mbox{\boldmath$r$}) = (\psi^{\prime}_{\alpha} +
i\psi_{\alpha}^{\prime \prime}); \alpha =1,.... n/2\}$ ($n=6$)
cannot be neglected~\cite{Uzunov:1993} and, as shown here, this
interaction produces new fluctuation phenomena.

In this letter a new critical behavior for this type of systems is
established and described. The new critical behaviour occurs in
real systems with isotropic magnetic order but does not belong to
any known universality class~\cite{Uzunov:1993}. Thus it could be
of considerable experimental and theoretical interest. Due to
crystal and magnetic anisotropy a new type of fluctuation-driven
first order phase transitions occur, as shown in the present
investigation. The quantum effects~\cite{Uzunov:1993, Hertz:1976,
Shopova:2003} on the properties of these novel phase transitions
are briefly discussed.

Both thermal fluctuations at finite temperatures ($T>0$) and
quantum fluctuations (correlations) near the $P$--driven quantum
phase transition at $T=0$ should be considered but at a first
stage the quantum effects~\cite{Shopova:2003} can be neglected as
irrelevant to finite temperature phase transitions ($T_F \sim T_c
>0$). The present treatment of a recently derived free energy
functional~\cite{Machida:2001, Walker:2002} by the standard
Wilson-Fisher renormalization group (RG)~\cite{Uzunov:1993} shows
that unconventional ferromagnetic superconductors with an
isotropic magnetic order ($m=3$) exhibit a very special
multi-critical behavior for any $T> 0$, whereas the magnetic
anisotropy ($m =1,2$) generates fluctuation-driven first order
transitions~\cite{Uzunov:1993}. Thus the phase transition
properties of spin-triplet ferromagnetic superconductors are
completely different from those predicted by mean field
theories~\cite{ShopovaPL:2003, Shopova:2005, Shopova:2006,
Machida:2001, Walker:2002}. The results can be used in the
interpretation of experimental data for phase transitions in
itinerant ferromagnetic compounds~\cite{Pfleiderer:2002}.

The study presents for the first time an example of complex
quantum criticality characterized by a double-rate quantum
critical dynamics. In the quantum limit ($T\rightarrow 0$) the
fields $\mbox{\boldmath$M$}$ and $\psi$ have different dynamical
exponents, $z_M$ and $z_{\psi}$, and this leads to two different
upper critical dimensions: $d_U^M = 6-z_M$ and $d_{\psi}^U =
6-z_{\psi}$. The complete consideration of the quantum
fluctuations of both fields $\mbox{\boldmath$M$}$ and $\psi$
requires a new RG approach in which one should either consider the
difference $(z_M-z_{\psi})$ as an auxiliary small parameter or
create a completely new theoretical paradigm of description. The
considered problem is quite general and presents a challenge to
the theory of quantum phase transitions~\cite{Shopova:2003}. The
obtained results can be applied to any natural system with the
same class of symmetry although this letter is based on the
example of itinerant ferromagnetic compounds.

The relevant part of the fluctuation Hamiltonian of unconventional
ferromagnetic superconductors~\cite{Shopova:2005, Shopova:2006,
Machida:2001, Walker:2002} can be written in the form
\begin{equation}
\label{eq1} {\cal{H}}= \sum_{\mbox{\boldmath$k$}}\left[ \left(r +
k^2 \right)|\psi(\mbox{\boldmath$k$})|^2 + \frac{1}{2}\left(t +
k^2\right)|\mbox{\boldmath$M$}(\mbox{\boldmath$k$})|^2   \right]
+\frac{ig}{\sqrt{V}}\sum_{\mbox{\boldmath$k$}_1,{\mbox{\boldmath$k$}_2}}{\mbox{\boldmath$M$}}
\left(
{\mbox{\boldmath$k$}}_1\right).\left[\psi\left({\mbox{\boldmath$k$}}_2
\right)\times \psi^{\ast}\left({\mbox{\boldmath$k$}}_1 +
{\mbox{\boldmath$k$}}_2 \right) \right]
\end{equation}
where $V \sim L^d$ is the volume of the $d-$dimensional system,
the length unit is chosen so that the wave vector
${\mbox{\boldmath$k$}}$ is confined below unity ($ 0 \leq k =
|{\mbox{\boldmath$k$}}| \leq 1)$, $g \geq 0$ is a coupling
constant, describing the effect the scalar product of
${\mbox{\boldmath$M$}}$ and the vector product
$(\psi\times\psi^{\ast})$ for symmetry indices $m = (n/2)=3$, and
the parameters $t \sim (T-T_f)$ and $r \sim (T-T_s)$ are expressed
by the critical temperatures of the generic ($g\equiv 0$)
ferromagnetic and superconducting transitions. As mean field
studies indicate~\cite{Machida:2001,Shopova:2003}, $T_s(P)$ is
much lower than $T_c(T)$ and $T_F(P) \neq T_f(P)$.

The fourth order terms ($M^4, |\psi|^4, M^2|\psi|^2$) in the total
free energy (effective Hamiltonian)~\cite{Shopova:2005,
Shopova:2006, Machida:2001, Walker:2002} have not been included in
eq.~(1) as they are irrelevant to the present investigation. The
simple dimensional analysis shows that the $g-$term in eq.~(1)
corresponds to a scaling factor $b^{3-d/2}$ and, hence, becomes
relevant below the upper borderline dimension $d_U=6$, while
fourth order terms are scaled by a factor $b^{4-d}$ as in the
usual $\phi^4-$theory and are relevant below $d<4$ ($b > 1$ is a
scaling number)~\cite{Uzunov:1993}. Therefore we should perform
the RG investigation in spatial dimensions $d = 6-\epsilon$ where
the $g$--term in eq.~(1) describes the only relevant fluctuation
interaction. Moreover, the total fluctuation
Hamiltonian~\cite{Shopova:2005, Shopova:2006, Machida:2001}
contains off-diagonal terms of the form
$k_ik_j\psi_{\alpha}\psi^{\ast}_{\beta}$; $i\neq j$ and/or $\alpha
\neq \beta$. Using a convenient loop expansion these terms can be
completely integrated out from the partition function to show that
they modify the parameters ($r,t,g$) of the theory but they do not
affect the structure of the model (1). Such terms change auxiliary
quantities, for example, the coordinates of the RG fixed points
(FPs) but they do not affect the main RG results for the stability
of the FPs and the values of the critical exponents. Here we
ignore these off-diagonal terms.

One may consider several cases: (i) uniaxial magnetic symmetry,
${\mbox{\boldmath$M$}} = (0,0,M_3)$, (ii) tetragonal crystal
symmetry when $\psi = (\psi_1,\psi_2,0)$, (iii) {\em XY} magnetic
order $(M_1,M_2,0)$, and (iv) the general case of cubic crystal
symmetry and isotropic magnetic order ($m=3$) when all components
of the three dimensional vectors ${\mbox{\boldmath$M$}}$ and
$\psi$ may have nonzero equilibrium and fluctuation components.
The latter case is of major interest to real systems where
fluctuations of all components of the fields are possible despite
the presence of spatial crystal and magnetic anisotropy that
nullifies some of the equilibrium field components. In one-loop
approximation, the RG analysis reveals different pictures for
anisotropic (i)-(iii) and isotropic (iv) systems. As usual, a
Gaussian (``trivial'') FP ($g^{\ast} =0$) exists for all $d>0$
and, as usual~\cite{Uzunov:1993}, this FP is stable for $d>6$
where the fluctuations are irrelevant. In the reminder of this
letter the attention will be focussed on spatial dimensions $d <
6$, where the critical behavior is usually governed by nontrivial
FPs ($g^{\ast} \neq 0$). In the cases (i)-(iii) only negative
(``unphysical''~\cite{Lawrie:1987}) FP values of $g^2$ have been
obtained for $d<6$. For example, in the case (i) the RG relation
for $g$ takes the form
\begin{equation}
\label{eq2} g^{\prime}= b^{3-d/2-\eta}g\left(1 +
g^2K_d{\mbox{ln}}b\right),
\end{equation}
where $g^{\prime}$ is the renormalized value of $g$, $\eta =
(K_{d-1}/8)g^2$ is the anomalous dimension (Fisher's
exponent)~\cite{Uzunov:1993} of the field $M_3$; $K_d =
2^{1-d}\pi^{-d/2}/\Gamma(d/2)$. Using eq.~(2) one obtains the FP
coordinate $(g^2)^{\ast} = -96\pi^3\epsilon$. For $d < 6$ this FP
is unphysical and does not describe any critical behavior. For $d
> 6$ the same FP is physical but unstable towards the parameter
$g$ as one may see from the positive value $y_g = -11\epsilon /2
>0$ of the respective stability exponent
$y_g$ defined by $\delta g^{\prime} = b^{y_g}\delta g$. Therefore,
a change of the order of the phase transition from second order in
mean-field approximation to a fluctuation-driven first order
transition when the fluctuation $g$--interaction is taken into
account, takes place. This conclusion is supported by general
concepts of RG theory~\cite{Uzunov:1993} and by the particular
property of these systems to exhibit first order phase
transitions~\cite{Shopova:2003} in mean field approximation for
broad variations of $T$ and $P$.

In the case (iv) of isotropic systems the RG equation for $g$ is
degenerate and the $\epsilon$-expansion breaks down. A similar
situation is known from the theory of disordered
systems~\cite{Lawrie:1987} but here the physical mechanism and
details of description are different. Namely for this degeneration
one should consider the RG equations up to the two-loop order. The
derivation of the two-loop terms in the RG equations is quite
nontrivial because of the special symmetry properties of the
interaction $g$-term in eq.~(1). For example, some diagrams with
opposite arrows of internal lines, as the couple shown in
fig.~(2), have opposite signs and compensate each other. The terms
bringing contributions to the $g$--vertex are shown
diagrammatically in fig.~3. The RG analysis is carried out by a
completely new $\epsilon^{1/4}$-expansion for the FP values and
$\epsilon^{1/2}$-expansion for the critical exponents; again
$\epsilon = (6-d)$. RG equations are quite lengthy and here only
the equation for $g$ is discussed. It has the form
\begin{equation}
\label{eq3} g^{\prime}=
b^{(\epsilon-2\eta_{\psi}-\eta_M)/2}g\left[1 + Ag^2 +
3(2B+C)g^4\right],
\end{equation}
where
\begin{equation}
\label{eq4} A=\frac{K_d}{2}\left[2{\mbox{ln}}b + \epsilon
({\mbox{ln}}b)^2 + (1-b^2)(2r+t)\right],
\end{equation}
\begin{equation}
\label{eq5} B= \frac{K_{d-1}K_d}{192}\left[9(b^2-1) -
11{\mbox{ln}}b - 6\left({\mbox{ln}}b\right)^2\right],\;\;\;\; C=
\frac{3K_{d-1}K_d}{64}\left[{\mbox{ln}}b +
2\left({\mbox{ln}}b\right)^2\right],
\end{equation}
$\eta_M$ and $\eta_{\psi}$ are the anomalous dimensions of the
fields ${\mbox{\boldmath$M$}}$ and $\psi$, respectively. The
one-loop approximation gives correct results to order
$\epsilon^{1/2}$ and the two-loop approximation brings such
results up to order $\epsilon$. In eq.~(4), $r$ and $t$ are small
expansion quantities with equal FP values $t^{\ast} = r^{\ast} =
K_d g^2$. Using the condition for invariance of the two
$k^2$-terms in eq.~(1) one obtains $\eta_M =\eta_{\psi} \equiv
\eta$, where
\begin{equation}
\label{eq6} \eta =\frac{K_{d-1}}{8}g^2\left(1-
\frac{13}{96}K_{d-1}g^2\right).
\end{equation}
Eq.~(3) yields a new FP
\begin{equation} \label{eq7}
g^{\ast}=8\left(3\pi^3\right)^{1/2}\left(2\epsilon/13\right)^{1/4},
\end{equation}
which corresponds to the critical exponent $\eta =
2(2\epsilon/13)^{1/2} - 2\epsilon/3$ (for $d=3$, $\eta \approx
-0.64 $).

The eigenvalue problem for the RG stability matrix
$\hat{{\cal{M}}} = \left[(\partial\mu_i/\partial
\mu_j);(\mu_1,\mu_2,\mu_3) = (r,t,g)\right]$ can be solved by the
expansion of the matrix elements up to order $\epsilon^{3/2}$.
When the eigenvalues $\lambda_j= A_j(b)b^{y_j}$ of
$\hat{{\cal{M}}}$ are calculated dangerous large terms of type
$b^2$ and $b^2(\mbox{ln}b)$, ($b \gg 1$)~\cite{Aharony:1974} in
the off-diagonal elements of the matrix $\hat{{\cal{M}}}$ ensure
the compensation of redundant large terms of the same type in the
diagonal elements $\hat{{\cal{M}}}_{ii}$. This compensation is
crucial for the validity of scaling for this type of critical
behavior. Such a problem does not appear in standard cases of RG
analysis~\cite{Uzunov:1993, Aharony:1974}. As in the usual
$\phi^4$--theory~\cite{Aharony:1974} the amplitudes $A_j$ depend
on the scaling factor $b$: $A_1=A_2=1+(27/13)b^2\epsilon$,
$A_3=1-(81/52)\epsilon(\mbox{ln}b)^2$. The critical exponents
$y_t=y_1$, $y_r=y_2$ and $y_g$ = $y_3$ are $b$--invariant:
\begin{equation}
\label{eq8} y_r = 2 + 10\sqrt{\frac{2\epsilon}{13}} +
\frac{197}{39}\epsilon,
\end{equation}
$y_t = y_r-18(2\epsilon/13)^{1/2}$, and $y_g = -\epsilon > 0$ for
$d < 6$. The correlation length critical exponents $\nu_{\psi} =
1/y_r$ and $\nu_M = 1/y_t$ corresponding to the fields $\psi$ and
${\mbox{\boldmath$M$}}$ are
\begin{equation}
\label{eq9} \nu_{\psi}= \frac{1}{2} -
\frac{5}{2}\sqrt{\frac{2\epsilon}{13}} +
\frac{103}{156}\epsilon,\;\;\;\;\; \nu_M = \frac{1}{2} +
2\sqrt{\frac{2\epsilon}{13}} - \frac{5\epsilon}{156}.
\end{equation}
These exponents describe a quite particular multi-critical
behavior which differs from the numerous examples known so far.
For $d = 3$, $\nu_{\psi}$ = 0.78 which is somewhat above the usual
value $\nu \sim 0.6 \div 0.7$ near a standard phase transition of
second order~\cite{Uzunov:1993},
 but $\nu_M = 1.76$ at the same dimension ($d=3$) is unusually large.
 The fact that the Fisher's
exponent~\cite{Uzunov:1993} $\eta$ is negative for $d=3$ does not
create troubles because such cases are known in complex systems,
for example, in conventional superconductors~\cite{Halperin:1974}.
The present $\epsilon$-expansion is valid under the conditions
$\epsilon^{1/2}b < 1$, $\epsilon^{1/2}(\mbox{ln}b) \ll 1$ provided
$b > 1$. These conditions are stronger than those corresponding to
the usual $\phi^4$-theory~\cite{Uzunov:1993,Aharony:1974}. This
means that the present expansion in non-integer powers of
$\epsilon$ has a more restricted domain of validity than the
standard $\epsilon$-expansion. Using the known
relation~\cite{Uzunov:1993} $\gamma = (2-\eta)\nu$, the
susceptibility exponents for $d=3$ take the values $\gamma_{\psi}
= 2.06$ and $\gamma_M = 4.65$. These values exceed even those
corresponding to the Hartree approximation~\cite{Uzunov:1993}
($\gamma = 2\nu = 2$ for $d=3$) and can be easily distinguished in
experiments.

The critical behavior discussed so far may occur in a close
vicinity of finite temperature multi-critical points ($T_c=T_f>0$)
in systems possessing the symmetry of the model (1). In certain
systems, as shown in Fig.~1, this multi-critical points may occur
at $T=0$. In the quantum limit ($T\rightarrow 0$), or, more
generally, in the low-temperature limit [$T \ll \mu; \mu\equiv
(t,r);k_B=1$] the thermal wavelengths of the fields
$\mbox{\boldmath$M$}$ and $\psi$ exceed the inter-particle
interaction radius and the quantum correlations fluctuations
become essential for the critical
behavior~\cite{Shopova:2003,Hertz:1976}. The quantum effects can
be considered by RG analysis of a comprehensively generalized
version of the model~(1), namely, the action ${\cal{S}}$ of the
referent quantum system. The generalized action is constructed
with the help of the substitution $(-{\cal{H}}/T) \rightarrow
S[{\mbox{\boldmath$M$}}(q),\psi(q)]$. Now the description is given
in terms of the (Bose) quantum fields $\mbox{\boldmath$M$}(q)$ and
$\psi(q)$ which depend on the $(d+1)$-dimensional vector $q =
(\omega_l, {\mbox{\boldmath$k$}})$; $\omega_l = 2\pi lT$ is the
Matsubara frequency ($\hbar=1;l = 0, \pm1,\dots$). The
${\mbox{\boldmath$k$}}$-sums in eq.~(1) should be substituted by
respective $q$-sums and the inverse bare correlation functions ($r
+ k^2$) and ($t + k^2$) in eq.~(1) contain additional
$\omega_l-$dependent terms, for example\cite{Hertz:1976,
Shopova:2003}
\begin{equation} \label{eq10}
\langle|\psi_{\alpha}(q)|^2\rangle^{-1} = |\omega_l|+ k^2 + r.
\end{equation}
The bare correlation function $\langle|M_j(q)|\rangle^2$ contains
a term of type $|\omega_l|/k^{\theta}$, where $\theta = 1$ and
$\theta =2$ for clean and dirty itinerant ferromagnets,
respectively~\cite{Hertz:1976}. The quantum dynamics of the field
$\psi$ is described by the bare value $z=2$ of the dynamical
critical exponent $z=z_{\psi}$ whereas the quantum dynamics of the
magnetization corresponds to $z_M = 3$ (for $\theta =1$), or, to
$z_M = 4$ (for $\theta = 2$). This means that the
classical-to-quantum dimensional crossover at $T\rightarrow 0$ is
given by $d \rightarrow (d + 2)$ and, hence, the system exhibits a
simple mean field behavior for $d \geq 4$. Just below the upper
quantum critical dimension $d_U^{(0)} =4$ the relevant quantum
effects at $T=0$ are represented by the field $\psi$ whereas the
quantum $(\omega_l$--) fluctuations of the magnetization are
relevant for $d < 3$ (clean systems), or, for even for $d < 2$
(dirty limit)~\cite{Hertz:1976}. This picture is confirmed by the
analysis of singularities of the relevant perturbation integrals.
Therefore, the quantum fluctuations of the field $\psi$ have a
dominating role below spatial dimensions $d <4$.

Taking into account the quantum fluctuations of the field $\psi$
and completely neglecting the $\omega_l$--dependence of the
magnetization ${\mbox{\boldmath$M$}}$, $\epsilon_0 =
(4-d)$--analysis of the generalized action ${\cal{S}}$ has been
performed within the one-loop approximation (order
$\epsilon_0^1$). In the classical limit ($r/T \ll 1$) one
re-derives the results already reported above together with an
essentially new result, namely, the value of the dynamical
exponent $z_{\psi}= 2 - (2\epsilon/13)^{1/2}$ which describes the
quantum dynamics of the field $\psi$. In the quantum limit ($r/T
\gg 1$, $T\rightarrow 0$) the static phase transition properties
are affected by the quantum fluctuations, in particular, in
isotropic systems ($n/2=m=3$). For this case, the one-loop RG
equations corresponding to $T = 0$ are not degenerate and give
definite results. The RG equation for $g$,
\begin{equation} \label{eq11} g^{\prime}=
b^{\epsilon_0/2}g\left(1 + \frac{g^2}{24\pi^3}\mbox{ln}b\right),
\end{equation}
yields two FPs: ({\em a}) a Gaussian FP ($g^{\ast} =0$), which is
unstable for $d<4$, and ({\em b}) a FP $(g^2)^{\ast} =
-12\pi^3\epsilon_0$ which is unphysical [$(g^2)^{\ast}<0$]  for
$d<4$ and unstable for $d \geq 4$. Thus the new stable critical
behavior corresponding to $T>0$ and $d<6$ disappears in the
quantum limit $T\rightarrow 0$. At the absolute zero and any
dimension $d > 0$ the $P-$driven phase transition (Fig.~1) is of
first order. This can be explained as a mere result of the limit
$T \rightarrow 0$. The only role of the quantum effects is the
creation of the new unphysical FP ({\em b}). In fact, the referent
classical system described by ${\cal{H}}$ from eq.~(1) also looses
its stable FP (7) in the zero-temperature ({\em classical}) limit
$T\rightarrow 0$ but does not generate any new FP because of the
lack of $g^3$--term in the equation for $g^{\prime}$; see
eq.~(11). At $T=0$ the classical system has a purely mean field
behavior~\cite{Shopova:2003} which is characterized by a Gaussian
FP ($g^{\ast} = 0$) and is unstable towards $T$--perturbations for
$0<d<6$. This is a usual classical zero temperature behavior where
the quantum correlations are ignored. For the standard $\phi^4-$
theory this picture holds for $d<4$. One may suppose that the
quantum fluctuations of the field $\psi$ are not enough to ensure
a stable quantum multi-critical behavior at $T_c=T_F=0$ and that
the lack of such behavior is in result of neglecting the quantum
fluctuations of $\mbox{\boldmath$M$}$. One may try to take into
account these quantum fluctuations by the special approaches from
the theory of disordered systems, where additional expansion
parameters are used to ensure the marginality of the fluctuating
modes at the same borderline dimension $d_U$ (see, e.g.,
Ref.~\cite{Shopova:2003}). It may be conjectured that the
techniques known from the theory of disordered systems with
extended impurities cannot be straightforwardly applied to the
present problem and, perhaps, a completely new supposition should
be introduced.

In conclusion, the present results may be of use in
interpretations of recent experiments~\cite{Pfleiderer:2002} in
UGe$_2$, where the magnetic order is uniaxial (Ising symmetry) and
the experimental data, in accord with the present consideration,
indicate that the C-P phase transition is of first order. Systems
with isotropic magnetic order are needed for an experimental test
of the new multi-critical behavior.

{\bf Acknowledgments.} The author thanks the hospitality of MPI-PKS
(Dresden) and ICTP (Triest) where a part of this research has been
performed. Financial support by grants No. P1507 (NFSR, Sofia) and
No. G5RT-CT-2002-05077 (EC, SCENET-2, Parma) is also acknowledged.

\begin{figure}
\epsfig{file=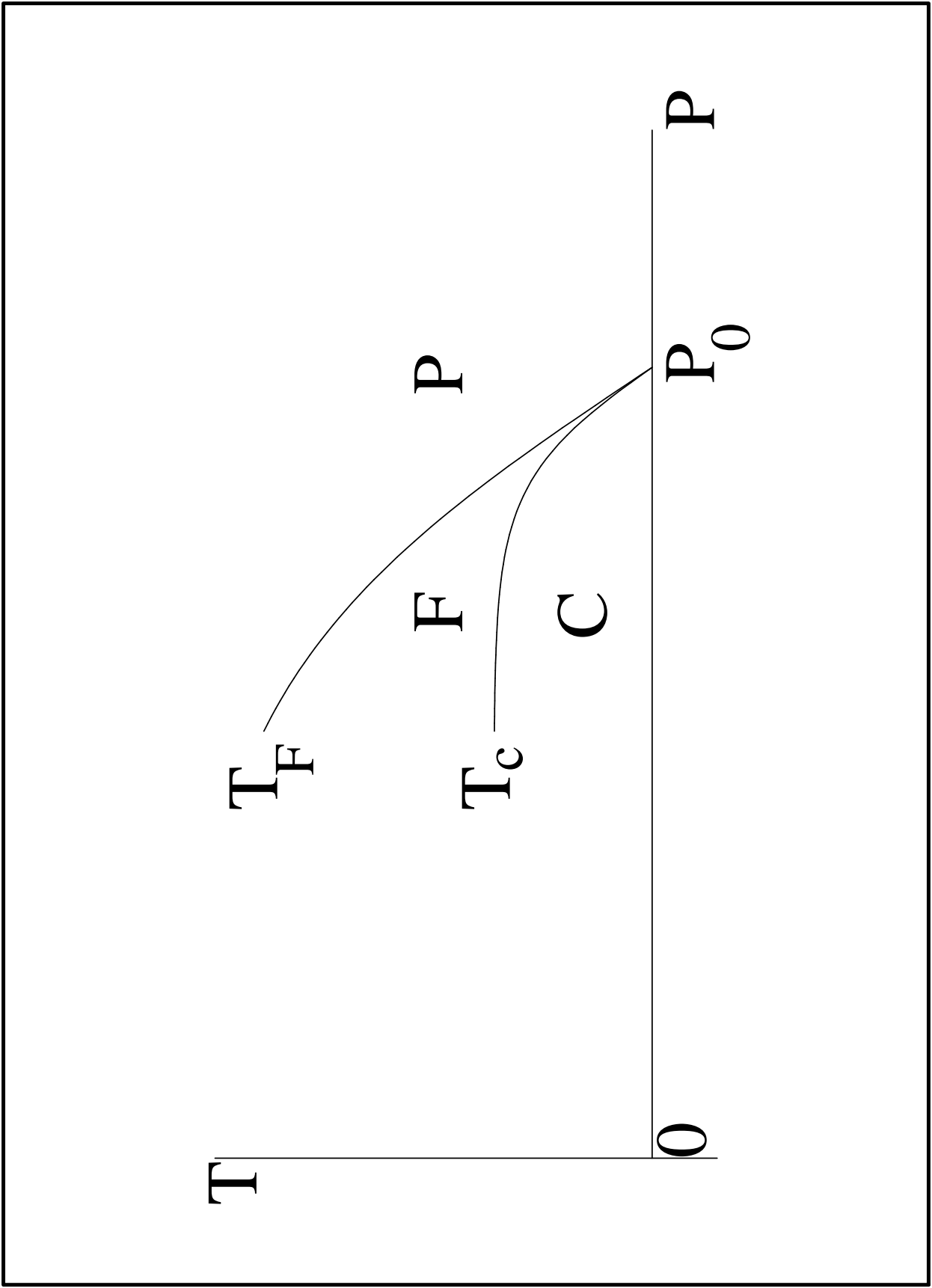}
 \caption{($P,T$) diagram with a
zero-temperature multicritical point $(P_0,0)$. Para- (P),
ferromagnetic (F), and coexistence (C) phases, separated by the
lines $T_f(P)$ and $T_c(P)$ of P-F and F-C phase transitions,
respectively.} 
\end{figure}

\begin{figure}
\epsfig{file=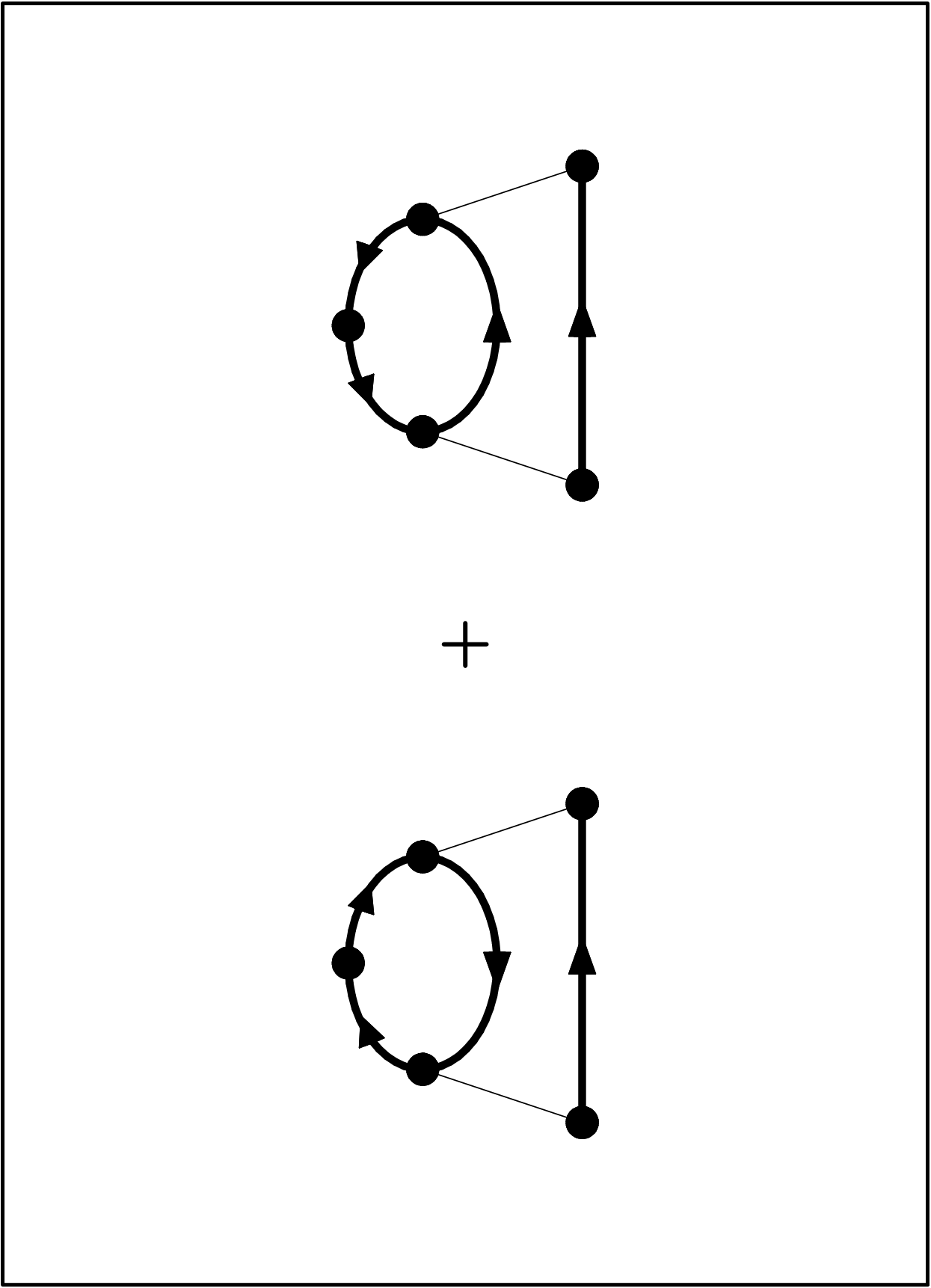} 
 \caption{A sum of $g^5$--diagrams equal to
zero. The thick and thin lines correspond to correlation functions
$\langle |\psi_{\alpha}|^2\rangle$ and $\langle |M_j|^2\rangle$,
respectively; vertices ($\bullet$) represent $g$--term in
eq.~(1).} 
\end{figure}

\begin{figure}
\epsfig{file=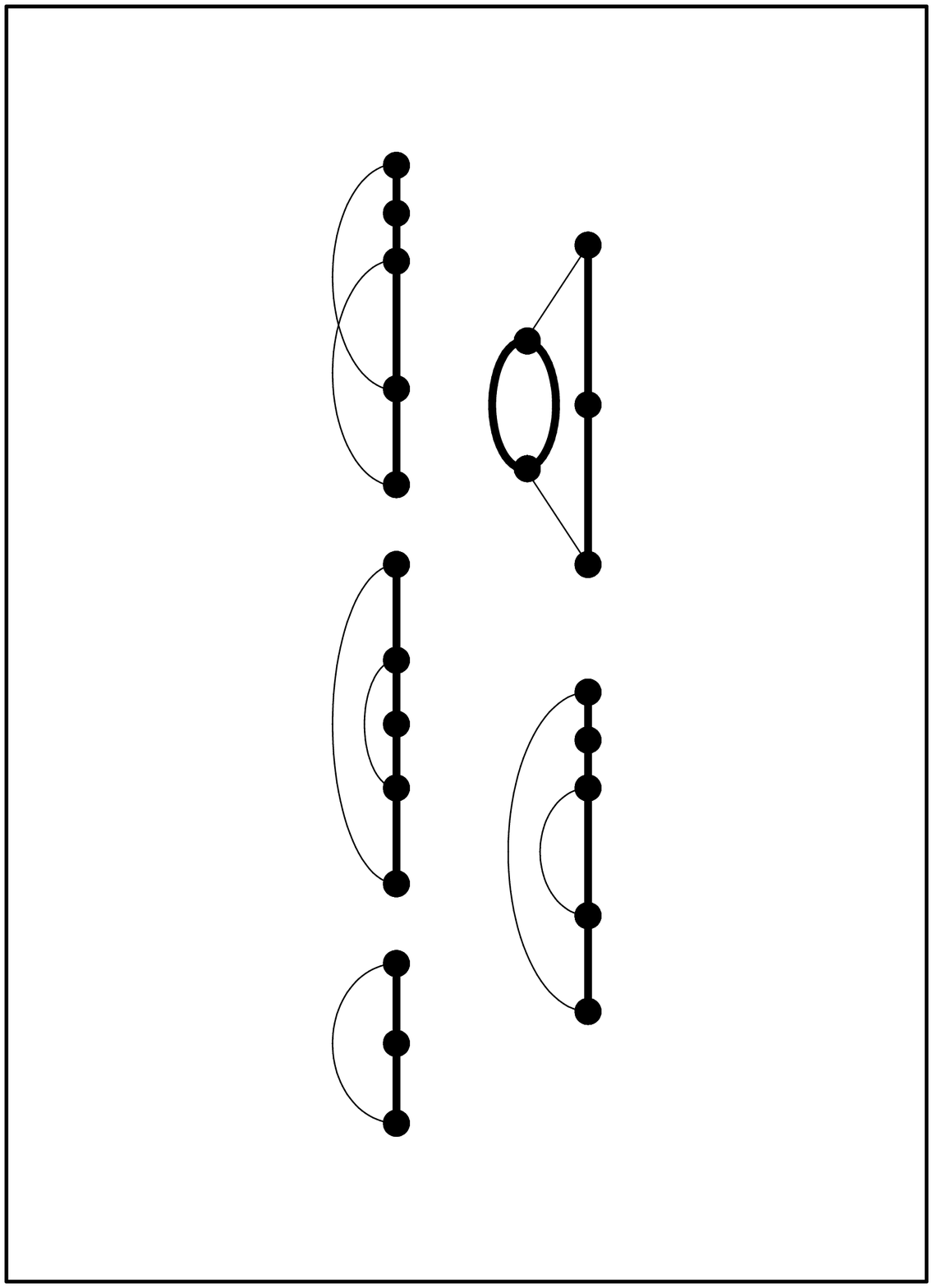} 
 \caption{Diagrams for $g^{\prime}$ of third
and fifth order in $g$. The arrows of the thick lines have been
omitted.}
\end{figure}


\begin{thebibliography}{ll}
\bibitem{Saxena:2000} Saxena S. S., Agarwal P., Ahilan K., Grosche F. M.,
Haselwimmer R. K. W., Steiner M. J., Pugh E., Walker I. R., Julian
S. R., Monthoux P., Lonzarich G. G., Huxley A., Sheikin I.,
Braithwaite D. and Flouquet J., Nature {\bf 406} {2000}
587.
\bibitem{Huxley:2001}
 Huxley A., Sheikin I., Ressouche E., Kernavanois N.,
Braithwaite D., Calemczuk R., and Flouquet J., Phys. Rev.
B {\bf 63} (2001) 144519.
\bibitem {ShopovaPL:2003} Shopova D. V. and D. I. Uzunov, Phys. Lett. A {\bf 
313} (2003) 139.
\bibitem {Shopova:2005} Shopova D. V. and D. I. Uzunov, Phys. Rev. B {\bf 72} 
(2005) 024531.
\bibitem {Shopova:2006}
Shopova D. V. and Uzunov D. V., in: Progress in Ferromagnetism Research,
 ed. bg V. N. Murrey
(Nova Science Publishers, New York, 2006) p.223.
\bibitem{Aoki:2001} Aoki D., Huxley A., Ressouche E., Braithwaite D., 
Flouquet J.,
Brison J-P., E. Lhotel and Paulsen C., Nature {\bf 413} (2001) 613.
\bibitem {Uzunov:1993}
Uzunov D. I., Theory of Critical Phenomena,
(World Scientific, Singapore, 1993).
\bibitem{Hertz:1976} Hertz J. A., Phys. Rev. B {\bf 14} (1976) 1165.
\bibitem {Shopova:2003} Shopova D. V. and Uzunov D. I., Phys. Rep. C {\bf 379}
 (2003) 1.
\bibitem {Sigrist:1991} M. Sigrist M. and Ueda K., Rev. Mod.
Phys. {\bf 63} 1991) 239.
\bibitem{Machida:2001} Machida K. and Ohmi T., Phys. Rev. Lett. {\bf 86}
(2001) 850.
\bibitem{Walker:2002} M. B. Walker M. B. 
and K. V. Samokhin K. V., Phys. Rev.
Lett. {\bf 88} 2002) 207001.
\bibitem{Pfleiderer:2002} Pfleiderer C. and Huxley A. D., Phys. Rev. Lett.{\bf
89} (2002) 147005.
\bibitem{Aharony:1974} Bruce A. D., Droz M. and Aharony A., J. Phys. C: Solid 
State Physics {\bf 7} (1974) 3673.
\bibitem {Lawrie:1987} Lawrie I. D., Millev Y. T. and D. I. Uzunov,
J. Phys. A: Math. Gen. {\bf 20} (1987) 1599.
\bibitem{Halperin:1974} Halperin B. I., Lubensky T. C. and
Ma S. K., Phys. Rev. Lett. {\bf 32} (1974) 292.
\end{thebibliography}
\end{document}